\documentclass[aps,prl,reprint,groupedaddress]{revtex4-1}
\usepackage{amsmath}
\usepackage{amsfonts}
\usepackage{amssymb}
\usepackage{graphicx}

\begin{document}

\title{Phase Transitions and Duality in Adiabatic Memristive Networks}


\author{Forrest C. Sheldon}
\email[]{fsheldon@ucsd.edu}

\author{Massimiliano Di Ventra}
\email[]{diventra@physics.ucsd.edu}
\affiliation{Department of Physics, University of California San Diego,
La Jolla, California 92093, USA}


\date{\today}

\begin{abstract}
The development of neuromorphic systems based on memristive elements - resistors with memory
- requires a fundamental understanding of their collective dynamics when organized in networks.
Here, we study an experimentally
inspired model of two-dimensional disordered memristive networks subject to a slowly
ramped voltage and show that they undergo a
first-order phase transition in the conductivity for sufficiently high
values of memory, as quantified by the memristive ON/OFF ratio. We investigate the consequences of this transition for the memristive current-voltage characteristics both through simulation and theory, and uncover a duality between forward and reverse switching processes that has also been observed in several experimental systems of this sort. Our work 
sheds considerable light on the statistical properties of memristive networks that are presently studied both for unconventional computing and as models of neural networks. 
\end{abstract}

\pacs{}

\maketitle


\section{Introduction}
Although systems that display resistive switching - also referred to as ``memristive elements'' (resistors with memory) - have been actively studied 
since the 1960s, they have recently received renewed interest in view of their possible use in computation, both as logic and memory components \cite{Pershin2011}.
Of particular note is the tendency of some to display a history-dependent decay constant, allowing a transition
between a volatile and non-volatile regime of memory \cite{Hasegawa2010, Ohno2011}.  The resulting dynamics
bear a close resemblance to the short-term and long-term potentiation observed in
biological synapses which are thought to be of central importance to learning and
plasticity in the brain~\cite{Chialvo2010}.  This resemblance has inspired the realization of experimental
systems that seek to combine the memory features of biological synapses with
the structural complexity of neural tissue \cite{Avizienis2012, Stieg2014}. In fact, research is being
performed to assess the advantage of using memristive elements in non-von
Neumann architectures and is already showing that their networks 
dynamically organize into the solutions of complex computational problems, thereby
performing the computation directly in the memory and avoiding the separation
between logic and memory units \cite{Ventra2013, Pershin2013, Traversa2015, Traversa15}.  

All of these studies however, still lack a fundamental understanding of the role of time non-locality in the dynamics of memristive networks and their statistical properties. For instance, high density 
($\sim 10^9$ elements/cm$^2$) disordered networks of memristive Ag/Ag$_2$S/Ag atomic
 switches have been fabricated showing collective switching behavior between a low-resistance ($G_{on}$) and high-resistance ($G_{off}$) state, and possible critical states potentially useful in neuromorphic computation \cite{Langton1990,Chialvo2010, Stieg2012}. Some theoretical work 
 has attempted to reproduce several features observed in the experiments by performing simulations in relatively small networks but an understanding of the dynamics of such systems is far from clear \cite{NedaaeeOskoee2011, Sillin2013}. 
Theoretical investigations of one-dimensional memristive networks have shown
complex temporal dynamics and scale-invariant properties, but
have not clarified whether further collective behavior might arise in
higher dimensions \cite{diVentra2013}.

To fill this gap, we study the statistical properties of two-dimensional disordered memristive networks subject to a slowly ramped voltage.  In this `adiabatic' regime, where the applied voltage/current varies much more slowly than the 
memristance change of individual elements, there is a strong analogy
between the behavior of the network and an equilibrium thermal system. Our aim is to understand the dynamics of both the disordered devices being
assembled in experiment, and the ordered (but strongly heterogeneous) networks being proposed as novel
computational architectures.  To this end, we formulate a general model for networks in this limit which is similar to those employed to describe metal-insulator transitions and electrical failure.  We thus posit that the transitions identified in these fields should also occur in memristive networks and summarize work done to describe the dynamics in these fields.  
Through simulations we obtain current-voltage (I-V) curves for various values of the $G_{on}/G_{off}$ ratio and discuss the features implied by the adiabatic limit.  The I-V curves found show a duality between forward and reverse switching processes that has been observed in several experimental systems \cite{Inoue2008, Sarswat2015} and clarifies the role of boundary conditions in the network. These features are captured by a mean-field theory and cluster approximation which clarify the internal dynamics and account for the features of the I-V curves. These results shed considerable light on the statistical and collective properties of networks with memristive elements that are being currently explored for neuromorphic applications.

\begin{figure*}
\includegraphics[width=17.8cm]{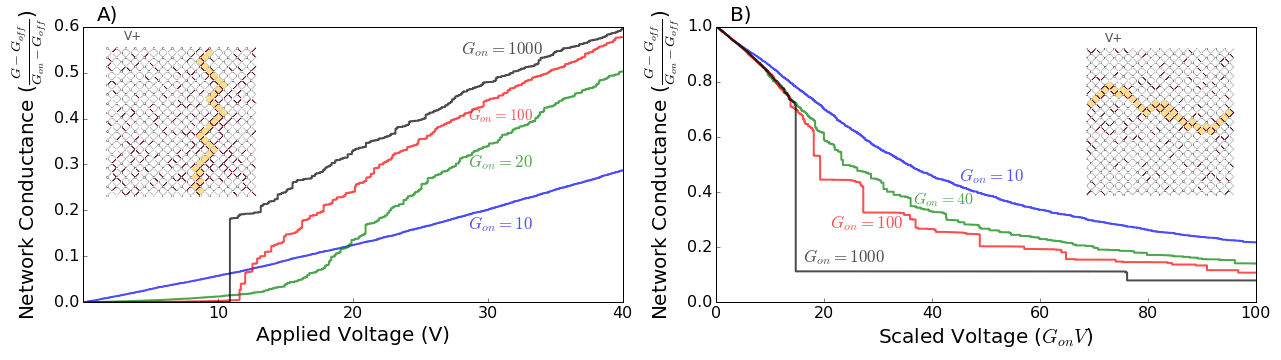}
\caption{The network conductance for several values of $G_{on}$ are plotted
against the applied voltage for both (A) $G_{off}\to G_{on}$ and (B) $G_{on}\to G_{off}$.
The network conductance $G$ has been scaled to
vary from 0 to 1 and the range of the applied voltage shortened to focus on
the point of transition. The insets show a typical network biased by a voltage $V_+$
just following the transition where the formation of a (A) conducting backbone and (B) crack
may be observed.
  \label{Cond_fig}}
\end{figure*}

\section{Model}

As inspiration and as a test bed for disordered memristive networks we consider
the Ag/Ag$_2$S/Ag gapless atomic switches experimentally fabricated in
\cite{Avizienis2012, Stieg2014, Sillin2014}. The model we consider though, is quite general and can be applied
to a variety of other materials and systems \cite{Pershin2011}.

A bipolar memristor subject to an external bias will transition to a conductive state (with conductance $G_{on}$)
in one direction and to an insulating state (with conductance $G_{off}$) when biased in the other.  This
change is generally induced by the rearrangement of ions in the applied electric field, as in the case of silver sulfide between two silver electrodes, where drifting ions form a filament structure eventually bridging the insulator \cite{Xu2010}.

These switching dynamics are typically subject to a threshold in the applied bias,
below which the conductance will not vary, or will vary only slowly.  When considering the dynamics of a single memristor, such thresholds may be described interchangeably in the applied voltage or current.  However when embedded in a network, the two can lead to quite different dynamics.  When the conductivity of an element within a network is increased, its current increases while its voltage decreases, and vice versa when its conductivity is decreased.  For the electric field driven migration of vacancies or ions~\cite{di2008electrical},  we expect that in order to simulate the behavior of actual devices, all elements must be subject to a current rather than voltage threshold.  The role of temperature has been emphasized by several studies, especially those focussing on the dissolution of the filament.  Such an effect may be taken into account by the explicit inclusion of a temperature in the model, or by taking a threshold in the power dissipated in an element $P_t = GI_t^2$ which gives a current threshold that changes with the device conductance $I_t = \sqrt{\frac{P_t}{G}}$.  In the discretely switching model we consider here, the two choices are identical, but when the full spectrum of memristances is allowed for, as in non-adiabatic regimes, such effects may be important.

The presence of a current threshold has immediate consequences on the dynamics of a network.  As the threshold of a device in the insulating state is crossed and its conductivity increases, more current is diverted through the element from neighboring bonds making the transition from $G_{off}$ to $G_{on}$ unstable.  On the timescale of the slowly varying applied voltage, elements of the network will appear to switch discretely between the insulating and conducting state. Therefore, we can
effectively model the elements as switching discretely from a conductance $G_{off}$ to a conductance 
$G_{on}$, when a threshold current, $I_t>0$, is crossed
\begin{equation}\label{gequation}
G(I) = G_{off} + (G_{on} - G_{off})\theta (I - I_t)\,,
\end{equation}
where $I$ is the current through the device, and $\theta(\cdot)$ is the Heaviside function of the argument. 

Similar considerations would lead us to conclude that the reverse direction is `stable', as passing an
element's threshold leads to a decrease
in the conductance and a corresponding decrease in the current, bringing it back below the threshold. The devices would thus explore the full continuum of memristance between $G_{on}$ and $G_{off}$ leading to a
gradual RESET-like behavior on the network level.  However, we find evidence both in experimental data and simulations that this effect does not occur or is not significant in describing the behavior of the network for a wide range of parameters.  For instance, in the atomic switch networks produced by Stieg {\it et al.} \cite{Stieg2012} sharp fluctuations in the network conductivity are observed in both directions, thought to be due to the switching of individual elements (see Fig. 3c in Ref. \cite{Stieg2012}).  Such sharp behavior may be accounted for by assuming a nonlinear dependence of the conductance on the filament length due to, for example, the transition from tunneling to ballistic conductance.  

Disorder at the
network level can similarly render the continuum of memristive values irrelevant to the network dynamics.
If the current diverted from a switching element does not cross another's threshold, the increasing current from the boundaries will continue to transition that element to $G_{off}$.  As the conductance and external voltage range in which this occurs is very small relative to the network scale, this is the same as if that element had switched discretely from $G_{on}$ to $G_{off}$.  Simulations of networks in which the full range of memristance was accessible have not shown a significant change in dynamics and we believe the discretely switching model to be appropriate for a wide range of networks.  
We thus make a similar assumption for the reverse direction, obtaining the equations for an element by exchanging $G_{on}$ with $G_{off}$, and
$I$ with $I_t$ in Eq.~(\ref{gequation}) for a different threshold $I_t<0$.

Memristive elements are also generally polar. The $Ag/Ag_2 S/Ag$ atomic switches
formed in atomic switch networks~\cite{Avizienis2012, Stieg2014} are gapless
type devices (see Hasegawa \textit{et al.} \cite{Hasegawa2012} for a
review of types of atomic switches and their switching processes).  Their symmetric metallic
configuration (typically gapless switches have two differing metallic
electrodes, e.g., $Ag/Ag_2 S/Pt$) suggests that at
the point of formation within the network, no 
preferred direction within the switch has been selected.  Polarity is instead instilled
through a `formation' step in which a bias is applied to the network
causing filament structures to form in the switches throughout. After
a joule-heating assisted dissolution of the thinnest part of the filament,
the junctions display bipolar resistive switching.  The polarity of the
internal switches is thus determined by the direction of current propagation
from the boundaries.  That this must be true in the experimental
systems is evident from the fact that the network undergoes resistive
switching as a whole.  Without a majority of switching polarities coinciding
with the direction of currents from the boundaries, the network would switch
between identical states with half the switches in the $G_{off}$ state and half in $G_{on}$
and not display the pinched hysteresis observed in experiment.  We thus assign the
polarity of elements in the network to coincide with the direction of currents flowing from the boundaries.

We now turn to a network of these elements.  We consider an 
architecture as depicted in the insets of Fig.~\ref{Cond_fig}, where the upper
and lower boundaries of the network are held to some constant 
voltage or total current. The diamond lattice is chosen such that all elements participate equally
in conduction and networks are periodic in the direction transverse to the current flow to mitigate
finite size effects.

Including disorder is most directly done through random pruning of the lattice, however the networks
produced are subject to strong finite size effects, requiring the simulation of many instances of large
networks to obtain regular results.  The random pruning of the lattice imposes a current
distribution over
the elements that is simply scaled by the external boundary conditions.
This distribution determines the order in which elements cross their thresholds and is thus equivalent, at a mean-field level, to assuming a distribution in the thresholds of a
fully occupied network.  It is also noted in~\cite{kahng1988electrical} that this is obtained upon coarse-graining a
randomly pruned lattice, and thus may be a better approximation to the thermodynamic limit than
simulations on structurally disordered lattices.  The use of a disorder distribution also affords us more
direct control over the relationship between the disorder and the dynamics, simplifying the search over a large
parameter space of possible structural disorders.


%


It is worth noting that this type of model has been arrived at in several contexts involving the interplay between conduction and disorder in
2D systems. For instance, a similar model was first applied to the study of the random fuse model for electrical failure ($G_{on} \to G_{off}=0$)~\cite{kahng1988electrical}. A uniform distribution of thresholds on the interval $[1-w,1+w]$ was considered and behavior examined as a function of network size $L$ and $w$.  Brittle and ductile regimes of behavior were identified, both of which culminated
in the formation of a lateral `crack' severing the network.  In the brittle, or narrow disorder regime, this occurred as an avalanche upon the first bond failing, while for larger $w$ there was a regime of diffuse failure, causing
the networks to progressively deform before the formation of a crack.  In the thermodynamic limit, only the brittle regime survived, except for the case $w\to 1$ when the disorder distribution extended to 0.  More recently, in
metal-insulator transitions (MIT) \cite{Shekhawat2011}  the $G_{off} \to G_{on}$ transition was examined for its $G_{on}/G_{off}$ dependence, finding that a transition occurs in which the network
conductivity exhibits a discontinuous jump, corresponding to the formation of a 'bolt'. Similarly, a conducting backbone has been found to form along the direction of current flow in the case of MIT and dielectric breakdown  \cite{Shekhawat2011} for sufficiently large values of the $G_{on}/G_{off}$ ratio.  

Interest in the dynamics of individual resistive switching (RS) devices has also led to new models such as the Random Circuit Breaker (RCB) model \cite{Chae2008} for unipolar devices, capable of reproducing the conductivity dynamics of a unipolar device in the SET and RESET operations.  In the RCB model, elements of a lattice transition to a conductive state when a voltage threshold is crossed, and back to an insulating state when another threshold is crossed in the same direction.

In view of these previous results, we thus expect the transitions observed in MIT and electrical failure studies to occur in memristive networks
in the adiabatic limit.  The $G_{off}\to G_{on}$ transition will correspond to the formation of a conductive backbone or `bolt' through the network along the direction of current flow and the $G_{on} \to G_{off}$ transition will correspond to the formation of a `crack' transverse to the direction of current flow severing the network.  In both directions we anticipate a trivial brittle, or narrow disorder regime in which the transition occurs upon the first element switching  and all elements transition within a narrow range of the applied voltage, as would be the case if all elements had the same threshold.  For broad disorder we expect a ductile regime in which there will be diffuse switching leading up to the transition and activity over a broad range of applied voltages.  While the ductile regime does not survive in the thermodynamic limit for electrical failure, the modest size
of memristive networks experimentally realized~\cite{Pershin2011} suggests
the ductile regime is still significant in their
dynamics. Of particular interest to us is the influence of such transitions on the I-V curves of the network.  From the perspective of the experimenter such effects may be desirable, such as providing strong sensitivity across a small voltage range or signaling the solution of a computational problem, or undesirable, by reducing the number of internal states accessible to the network. 

Investigations of the behavior of memristive networks have been limited to date.  In \cite{NedaaeeOskoee2011}, the full time integration of a memristive network was undertaken for networks of moderate size.  Elements did not include a threshold in their dynamics and the networks were investigated for their dependence on the fraction of memristive to ohmic conductors $p$, and their AC response.  It was found that for poor ohmic conductors ($G=G_{off}$) the networks exhibited pinched hysteresis curves only when $p>0.5$, the percolation threshold.  For good ohmic conductors ($G=G_{on}$), a strongly memristive phase was observed for $p>p_c$ in which the networks switched abruptly and for $p<p_c$ a weakly memristive phase was observed, similar to that for poor ohmic conductors and $p>p_c$.  

Modeling performed by \cite{Sillin2013} of the ``atomic switch networks''
simulated small networks including volatility and noise in their memristor model, and showed an opening of the I-V curves as the noise term was reduced and $1/f^\gamma$ ($0\le \gamma \le 2$) scaling of the power spectral density for the small networks simulated ($L\approx 10$). While such studies reproduce phenomena seen in experiment, little work has been done to analyze the behavior of these models and understand how memristors in networks interact.  Here, we instead focus on building an understanding of memristive networks in the adiabatic regime, where analogies with thermodynamic systems are strongest.  By moderating the strength of interactions through the $G_{on}/G_{off}$ ratio, we examine the transition that occurs in each direction and it's effects on the I-V curves of the network.  The features of this transition and the resulting hysteresis curves are well captured by a cluster approximation that well approximates the behavior of the network about the point of transition.

\begin{figure*}
\includegraphics[width=17.8cm]{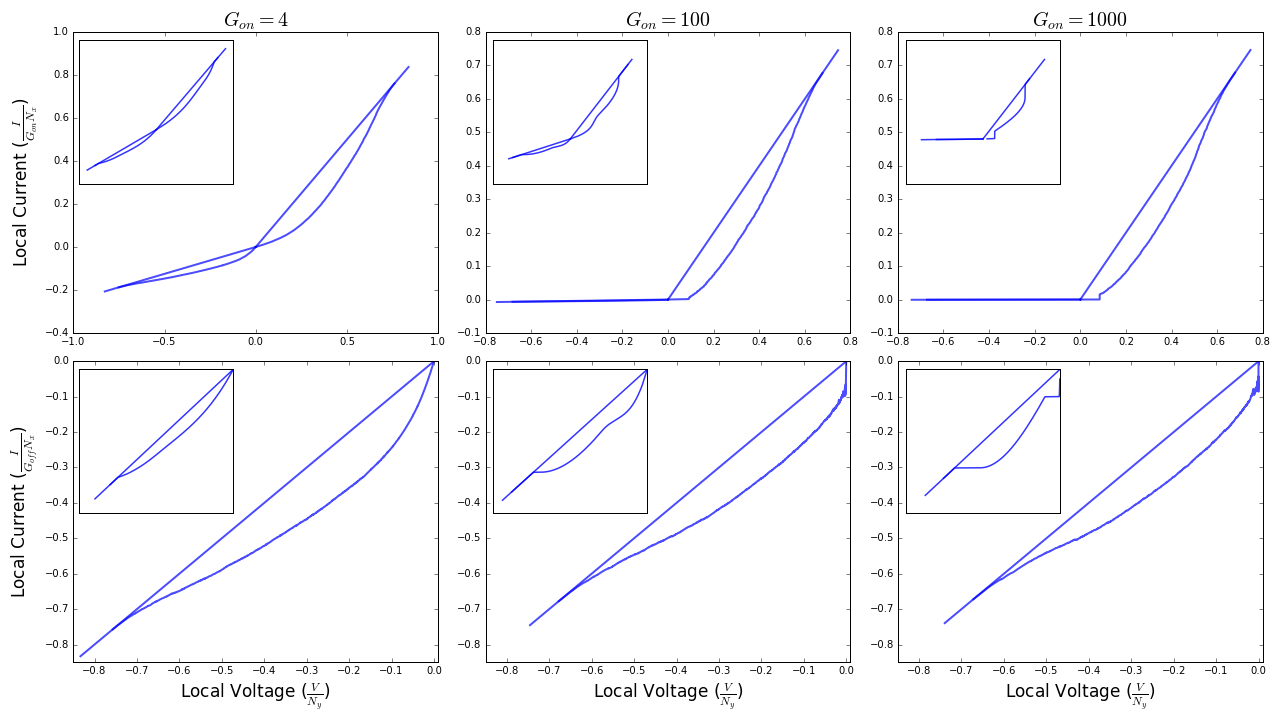}
\caption{Simulated hysteresis curves of memristive networks for $G_{on} = $ 4, 100, and 1000.  The discrete jump in conductance becomes evident in the I-V curves for large values of $G_{on}$
The asymmetry of the curves is the result of element thresholds being in terms of the current, and thus
the transition occurring at a factor of $1/G_{on}$ lower voltage than the corresponding forward transition.
The insets show analytically reached I-V curves which demonstrate similar asymmetry and the emergence of a jump in the current.  The reverse branch has been rescaled and plotted in the second row.  For large $G_{on}$, the transition appears as a noisy area near to the y-axis.  Here the jumps in conductivity at a fixed voltage give rise to sharp decreases
in the current
that are opposed by the subsequent increase in voltage.
  \label{IV_fig}}
\end{figure*}

\section{Simulations}
Simulations were carried out for a square lattice at a
variety of sizes,
$G_{on}/G_{off}$ ratios, and threshold distributions $p(t)$.  The network dimensions
were chosen such that the conductivity of the network varied from $G_{off}$
to $G_{on}$.  Each element was
assigned a current threshold $I_t$ from $\text{Uniform}(0,1)$  in each direction, beyond which they transition from $G_{off}\to G_{on}$, or vice versa.  Network dimensions were chosen so that the total network conductance varied between $G_{off}$ and $G_{on}$  ($N_x = N_y = 128$).  The initial voltage is set to the value required to cross the lowest threshold in the network.  Once that element has switched, voltages and currents are recalculated throughout the network with the external voltage held fixed, and all other elements whose currents exceed their thresholds
are switched.  This is repeated until no currents exceed the thresholds of their elements, at which point the voltage is raised until another threshold is crossed and this process repeated.  The forward and reverse protocols are identical aside from the initial state and switching direction of the elements.

In Fig.~\ref{Cond_fig} we show the network conductances as a function of
applied voltage for various values of $G_{on}$ (setting $G_{off} = 1$), in both forward
$G_{off}\to G_{on}$  and backward $G_{on}\to G_{off}$ transitions, and for
threshold distribution $p(t) = \textrm{Uniform}(0,1)$.  The displayed networks have
a linear size of $N_{x/y}=128$ memristors which we found large enough to achieve
regular results over multiple realizations of the disorder. Network conductances
have been scaled to lie on the interval $[0,1]$. We note that for small values
of $G_{on}$  in both directions, the conductance is a smooth function of the voltage. 
As $G_{on}$
is increased, a discontinuity forms in the slope which sharpens, appearing
almost continuous until a discontinuous jump appears for large $G_{on}$ analogous to a {\it first-order phase transition}.  Similar behavior was seen for
a variety of other distributions of sufficient breadth (not shown) with the point of transition, however, being
distribution dependent. In the insulating transition, we have scaled the voltage by $G_{on}$
such that the current densities of all networks are initially equal, bringing the transitions to the same scale in both polarities.  While the dependence on $G_{on}/G_{off}$ in the forward direction has been shown
 in MIT \cite{Shekhawat2011}, we are not aware of a similar demonstration in the reverse direction, possibly as most work has focused on electrical breakdown in fuse networks in which $G_{off} = 0$.
 
 The insets in Fig.~\ref{Cond_fig} each show the configuration of a network just following a transition.  In
 the forward transition (A), this corresponds to the emergence of a conducting backbone spanning the network.
 Similarly, in the reverse transition (B) a crack forms separating the network transverse to the direction
 of current flow.

 \begin{figure*}
\includegraphics[width=17.8cm]{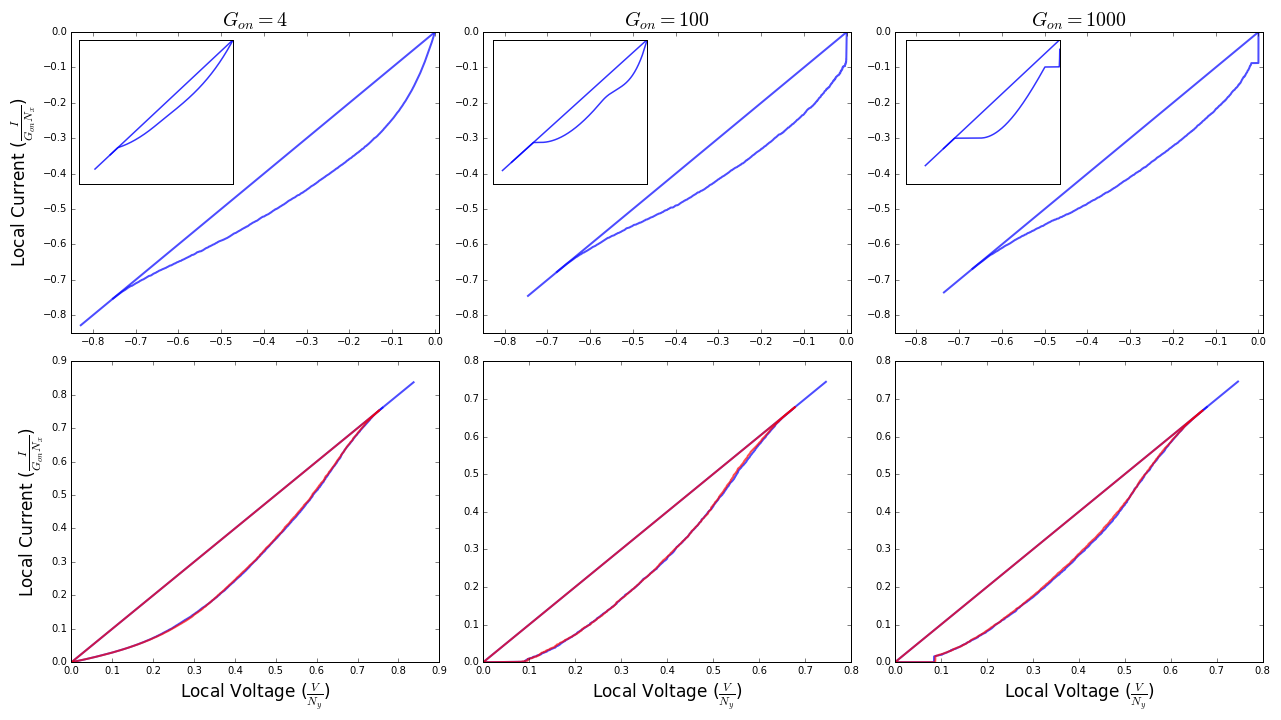}
 \caption{For a current-controlled network switching from $G_{on}\to G_{off}$ the transition gives a
 discrete jump, as in the voltage-controlled $G_{off}\to G_{on}$ case.
 After the mapping in Eqn.~\ref{BF_mapping_eqn},  the reverse branch of the hysteresis curves (red)
 have been plotted over the forward branch for several values of $G_{on}$ demonstrating the duality between the two processes.
 \label{backward_forward}}
 \end{figure*}
 
The corresponding I-V curves are shown in Fig.~\ref{IV_fig}.   While the voltage scale depends only on the geometry of the network, the current scale depends on $G_{on}/G_{off}$ and has been rescaled so that the axes coincide.  The strong asymmetry of the curves is due to the use of the same threshold distribution for $G_{on}\to G_{off}$ and $G_{off}\to G_{on}$ processes.  In the $G_{on}$ state, an equivalent current will be reached for a voltage which is a factor $1 / G_{on}$ lower.  As there is not an obvious physical choice for how to connect the distributions for forward and reverse switching, we have displayed the negative voltage sections of the curves separately, on their own current scales.    The discrete jump in the conductance becomes evident for sufficiently large values of $G_{on}$, but is less apparent than in the conductivity plots due to the long tail following  the transition.  Past the transition, voltage steps between thresholds become longer as current is diverted into the
conducting backbone.  The transition thus has an inhibitory effect on the remaining bonds, opening the hysteresis curves and spreading the memristive states over a wider voltage range.  Thus, while the transition reduces the number of accessible states, it increases the resolution between those remaining.  

In the second row of Fig. ~\ref{IV_fig}, the scaled reverse branches of the I-V curves appears remarkably similar to the positive branch, but with the roles of the current and voltage exchanged.  This suggests the following mapping of the reverse branch to the forward branch,
\begin{equation}
V\to \frac{IN_y}{N_x G_{on}}, \quad I \to \frac{VG_{off} N_x}{N_y},
\label{BF_mapping_eqn}
\end{equation}
where $N_x$ and $N_y$ are the lattice dimensions.
In the region about the transition however, the jump of the forward branch appears as a fluctuating region in the reverse branch.  Here, avalanches of transitioning elements sharply reduce the current, which is then opposed by a subsequent increase of the externally applied voltage.  If we instead run the reverse branch with current-controlled boundary conditions, this fluctuating region becomes a discrete jump as seen in the first row of Figure~\ref{backward_forward}. In the second row, upon the mapping (\ref{BF_mapping_eqn}), the curves align nearly exactly.
This correspondence between the two processes is just the familiar I-V {\it duality} of electrical circuit theory \cite{Iyer1985}: the diamond
lattice is dual to itself and taking $G\to \frac{1}{G}$ in all links takes a voltage-controlled insulator-to-metal
transition to a current-controlled electrical failure process.  

In the context of memristor networks, this indicates that the voltage-controlled I-V curves are dual to the current-controlled I-V curves upon exchanging the roles of voltage and current \emph{and} the direction of the switching process. Running the model in the current controlled setting thus gives
nearly identical results, but exchanges the fluctuating region in the reverse direction for the discrete jump in the forward direction.  

It is important to note that this connection between the forward and reversed hysteresis loops has been observed experimentally in individual memristive systems \cite{Inoue2008, Sarswat2015} as well.  While observing the duality for a network of resistors is clear, the extension to the dynamic non-linear elements considered here is surprising, especially when considering the asymmetry in the switching process between the forward and reversed directions, one of which corresponds to a backbone forming along the direction of current propagation, and the other corresponding to a crack forming transverse to the current flow.  

In the remainder of the paper, we analytically investigate our model with the aim of understanding the major features of the simulated I-V curves, namely the existence of a transition for sufficiently large $G_{on}/G_{off}$, the long tail following the transition, and the duality between the forward and reverse switching processes, leading to the I-V curves displayed in the insets of Figs. \ref{IV_fig} and \ref{backward_forward}.

\section{Mean-Field Theory}

As a first step towards an analysis of the model, we develop its mean-field theory. The method followed is similar
to that of Zapperi {\it et al.}
\cite{Zapperi1999} employed to analyze random fuse networks.  In this form, the central physical quantity considered
is the power dissipated by the network,
\begin{equation}\label{power}
	P = \sum_j g_j v_j^2 = \sum_j \frac{i_j^2}{g_j}
\end{equation}
where $g_j$ is the conductance of an element in the network and $v_j (i_j)$ is its voltage drop (current).   We require that the average power dissipated match the power dissipated by the network $G_{net}V^2 = \frac{I^2}{G_{net}}$ and assume that all elements experience a mean-field voltage $V_{MF}$ or current $I_{MF}$ leading to the equations,
\begin{equation}
\left. \begin{aligned}
G_{net} V^2 \\
\frac{I^2}{G_{net}}
\end{aligned} \right\} = 
\begin{cases}
N\langle g\rangle V_{MF}^2 \\
N \langle \frac{1}{g}\rangle I_{MF}^2.
\end{cases}
\label{MF_eqns}
\end{equation}
The choice of the LHS is determined by the boundary conditions applied to the network but the choice of the RHS is not constrained.  An interesting form is the `voltage-voltage' choice, leading to the mean-field voltage
\begin{equation}
V_{MF} = \sqrt{\frac{G_{net}}{\langle g \rangle}}\frac{V}{\sqrt{N}}
\end{equation}
which unlike other choices displays a transition in both directions.  To make progress we require
the network conductance $G_{net}$.
Below the transition,
where switching of elements is primarily driven by the threshold distribution and not by the influence of nearby switched elements, the conductivity of the network may be well approximated by an effective medium theory, giving $G_{net} \approx G_{eff}(f)$ as a function only of the fraction of the devices in the ON state.

\begin{figure}
\includegraphics[width=8.6cm]{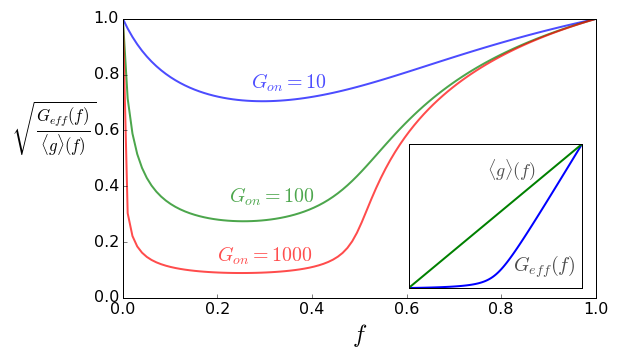}
\caption{ $h(f) = \sqrt{\frac{G(f)}{\langle g\rangle(f)}}$ is plotted for several values of
$G_{on}$.  In the inset, $\langle g \rangle(f)$ and $G(f)$ are plotted for $G_{on}=100$.
Note that when the average conductance increases more quickly than the network
conductance as for small $f$, $h(f)$ is decreasing and vice versa
for large $f$.\label{MF_voltage_fig}}
\end{figure}

The functions $\langle g\rangle(f)$, $G_{eff}(f)$,
and $h(f) = \sqrt{\frac{G_{eff}(f)}{\langle g\rangle (f)}}$ are plotted in Figure \ref{MF_voltage_fig}. 
We can understand the non-monotonic
form of $h(f)$ as arising from competition between switching elements
concentrating current away from other elements, and the increasing conductance
of the network pulling more current in at the boundaries.  For small $f$, the
average conductance of an element is increasing faster than the network
conductivity, indicating that current is concentrated away from other elements
more quickly than it increases at the boundaries, and the mean-field voltage
decreases.  For larger $f$, the network conductance begins to increase more
quickly than the average conductance, pulling in current faster than switching
elements can concentrate it, and the mean-field voltage increases.  The
increasing regime at large $f$ allows for a phase transition to occur from
$G_{off}\to G_{on}$, and the decreasing portion at low $f$ allows for the possibility
of the reverse transition when the voltage is reversed.  

\begin{figure}
\includegraphics[width=8.6cm]{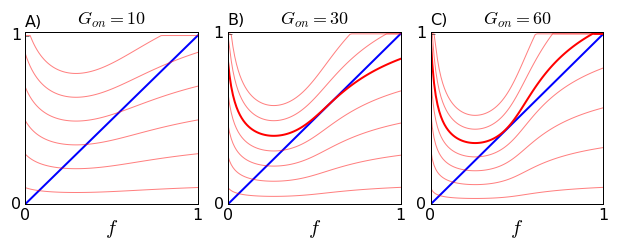}
\caption{The lhs (blue) and rhs (red) of Eq.~(\ref{selfconsist}) are plotted
for several values of the applied voltage.  For low values of $G_{on}$ their
intersection gives a solution that is smooth function of the voltage (panel A).
A transition develops for intermediate values that can appear continuous
(panel B). For sufficiently large values of $G_{on}$, a transition occurs where the
solution jumps discontinuously.  The transition voltage is highlighted in
panels B) and C).
\label{MF_SC_fig}}
\end{figure}

To determine whether a transition occurs for a particular disorder
distribution, we derive a
self-consistency equation, ensemble-averaging over the number of elements
that have switched for a given mean-field voltage. For the transition from
$G_{off}$ to $G_{on}$, the fraction that has switched will approach the average fraction
of elements with thresholds below the mean-field voltage,
\begin{equation}\label{selfconsist}
f = \int_0^{h(f) v} p(t) dt.
\end{equation}
Because the applied field enters multiplicatively, the dynamics given by
the mean-field theory depend only on the conductance ratio $G_{on}/G_{off}$
and is independent of the length scale of the disorder, both amounting only to
a rescaling of the applied local field $v$.  The lhs and
rhs of Eq.~(\ref{selfconsist}) are plotted for several values of the voltage
in Figure \ref{MF_SC_fig}.  For
the chosen distribution and $G_{on}/G_{off}$ ratios a transition is evident at the
point
\begin{equation}\label{PT_eqn}
1 = p(h(f)v)h'(f)v \quad 0\le f\le 1.
\end{equation}

We also note that the inflection
of the curves from the RHS of Eq.~(\ref{selfconsist}) shows a trend that looks almost like a continuous
transition, corresponding to the behavior seen in simulations for intermediate
values of $G_{on}$ (see Figure \ref{Cond_fig}).

An exactly analogous treatment may be undertaken for the transition from $G_{on} \to G_{off}$. 
We regard $f_R=1-f$ as the fraction of devices in their $G_{off}$ state and $v=\frac{V}{\sqrt{N}}$ as the positive
voltage applied in the \emph{reverse} direction.  The effective medium conductivity may be obtained
from the substitution $f\to 1-f_R$, 
$G_{eff,R}(f_R) = G_{eff}(1-f_R)$ and similarly with the average conductivity $\langle g\rangle_R(f_R) = \langle g\rangle (1-f_R)$, giving
a mean field voltage $V_{MF} = h_R(f_R)v = \sqrt{\frac{G_{eff,R}(f_R)}{\langle g\rangle_R(f_R)}}v$.  As the mechanisms for
turning ON and OFF within the individual atomic switches are not the same, we take a possibly different
probability distribution $p_R(t)$ for the reverse switching thresholds.  With these definitions we obtain the
self-consistency equation as before,

\begin{equation}
f_R = \int_0^{h_R(f_R)v} p_R(t) dt.
\label{selfconsist_R}
\end{equation}
In Fig. \ref{MF_SC_fig} this quantity has been plotted for several values of $G_{on}$ and the applied voltage. We observe a first-order phase transition similar to that observed in the
$G_{off}\to G_{on}$ branch. However, this transition occurs for much lower values of $G_{on}$ and near the limiting
value of the conductance.  As we proceed from $f=1$ to $0$, in the vicinity of $f=0$ the average conductance $\langle g\rangle(f)$ is decreasing more rapidly than the network conductance $G_{eff}(f)$.  This leads the internal switches to redistribute current to their neighbors faster than the decrease of the total current at the boundary.  This increases the mean-field voltage overall, and hence promotes a transition.

\begin{figure}
	\includegraphics[width=8.6cm]{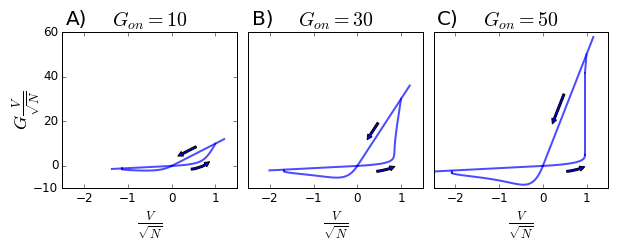}
	\caption{Solving the self consistency equations (\ref{selfconsist}) and (\ref{selfconsist_R}) for a uniform distribution leads to the hysteresis curves above.  For small values of $G_{on}/G_{off}$, the networks are smooth, but as the ratio is increased a discrete jump emerges in the forward direction.  A similar jump near the end of the reverse branch is only barely discernible.
		\label{MF_hysteresis}}
\end{figure}

Solving the self-consistency equations gives the mean-field hysteresis curves plotted in  Fig.~\ref{MF_hysteresis}.  While these curves display a qualitative similarity to simulation, several features of the mean-field theory are lacking.  As we have already noted, the choice for the form of the mean-field theory  (voltage-voltage, current-voltage, etc. ) are not prescribed {\it a priori} and each choice will give a slightly different account of the dynamics. All of these share the feature that  transitions occur due to competition between a changing current at the boundaries and the internal sharing of currents within the network, as summarized by the function $h(f)$ which is some ratio between the network conductance and an average conductance $\langle g\rangle, \,\langle \frac{1}{g}\rangle$.  In both directions, the transition will eventually proceed (as $G_{on}\to\infty$) from some critical fraction $f_c$ to a fully switched network $f=1$ as opposed to the finite jump and long tail seen in simulations.  

Having observed the internal form of the transition in simulation, we can see that in contrast to thermal transitions, where the phase transition occurs homogenously throughout the system, the conductivity transitions consist only of a $d=1$ dimensional conducting backbone in the forward direction and a $d=D-1$ dimensional crack in the reverse.  In $D=2$ both of these correspond to one dimensional subsystems of the network and so the mean-field theory which considers all elements equally cannot model it accurately, especially in the regime following the formation of the backbone. In the following, we consider methods for modeling the formation of the backbone.

\section{1D Models}
The mean-field assumption, that all elements experience either a voltage $V_{MF}$ or current $I_{MF}$,
is equivalent to replacing the network with a 1D parallel or series arrangement of memristors whose boundary conditions are then matched (through an effective medium theory) to the behavior of the original network.  In fact, the quantities
\begin{equation}
\langle g \rangle = G_{off} + \frac{n}{N}(G_{on} - G_{off})
\end{equation}
\begin{equation}
\quad \left\langle \frac{1}{g} \right\rangle = \frac{G_{on} +  \frac{n}{N} (G_{off} - G_{on})}{G_{on}G_{off}}
\end{equation}
which appear in the mean-field equations~(\ref{MF_eqns}) are also the conductances $G_{net}/N$ 
for a series and parallel network of $N$ memristors with $n$ in the ON state.  Having seen that the transition is restricted to a small subset of the network, we do not expect that including the entire network in the backbone will capture the behavior in the vicinity of the transition (where homogeneity, and thus the effective medium theory, fails).  We thus first explore the opposite extreme by ignoring the presence of the rest of the network and considering only those elements involved in the conducting backbone or crack.

\begin{figure}
	\includegraphics[width=8.6cm]{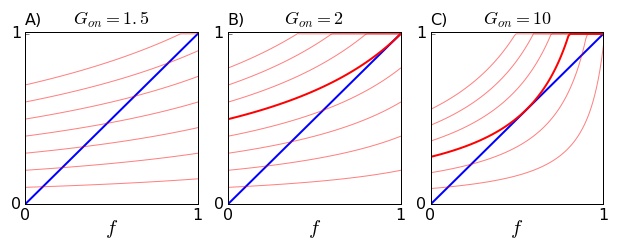}
	\caption{The lhs (blue) and rhs (red) of Eq.~(\ref{SC_1D_Forward_eqn}) are plotted
		for several values of the applied voltage for the distribution $\text{Uniform}(0,1)$.  For a
		collection of memristors in series, the transition is the completion of the conducting backbone,
		with all elements in the $G_{on}$ state.
		\label{SC_1D_Forward}}
\end{figure}

In the forward direction, the transitioning elements are a collection of memristors in series of length $N_y$ held at a voltage $V$.  Such an arrangement with a fraction $f$ in the ON state admits a current,
\begin{equation}
I(f) = \frac{G_{on}G_{off}}{G_{on} +  f (G_{off} - G_{on})} \frac{V}{N}
\end{equation}
and the fraction of elements in the ON state may be determined self-consistently,
\begin{equation} 
f = \int_0^{I(f)} \rho(t) dt.
\label{SC_1D_Forward_eqn}
\end{equation}
The distribution $\rho(t)$ is the distribution of thresholds in the conducting backbone which should be related to the distribution of thresholds across the network.  While an explicit calculation of $\rho$ is difficult, a reasonable approximation on the diamond lattice should be $\rho(t) = 2p(t)(1-F(t))$ where $F(t)$ is the cumulative distribution function of the threshold distribution, such that the current always selects the path with lower threshold. We note that this concentrates the threshold distribution towards its lowest values but does not strongly alter the behavior of the theory.  In the interest of simplicity, we thus maintain our use of the $\text{Uniform}(0,1)$ distribution in illustrating the features of each approach. Equation~(\ref{SC_1D_Forward_eqn}) is plotted in Fig.~\ref{SC_1D_Forward} for several values of $G_{on}$ in which we observe the transition first occurring at  $G_{on}=2$, and then progressing to a jump to $f=1$ for larger values.

In the reverse direction, we consider a collection of memristors in parallel of length $N_x$ corresponding to the crack that will eventually sever the network.  As the `crack' is separated from the boundaries, the boundary conditions are instead supplied by the network.  As every strip of memristors perpendicular to the direction of current propagation will have a current $G_{net} V$ passing through them, the reverse switching process of a voltage-controlled network should be best described by a current-controlled strip of memristors in parallel.  For the moment, we again ignore the presence of the rest of the network and consider an isolated set of elements.  The conductance of the 1D strip of memristors in parallel with a fraction $f_R$ in the $G_{off}$ state is
\begin{equation}
N(G_{on} + f_R(G_{off} - G_{on}))
\end{equation}
which gives the current through an element in the ON state,
\begin{equation}
I(f_R) = \frac{G_{on} G_{net}}{G_{on} + f_R(G_{off} - G_{on})}\frac{V}{N}.
\end{equation}
$f_R$ may be similarly found self-consistently
\begin{equation}
f_R = \int_0^{I(f_R)} \rho(t) dt.
\end{equation}
The resulting mean-field theory is just the reverse of that for the forward switching process (taking $G_{net} = G_{on}$) but with a voltage scale smaller by a factor of $G_{on}$.

Here, physical considerations from the switching processes have led us to two {\it dual} structures: a series chain of memristors transitioning from $G_{off}\to G_{on}$ subject to a ramped voltage as a model of the conducting backbone, and a parallel strip of memristors transitioning from $G_{on}\to G_{off}$ subject to a ramped current for the crack severing the network.  Each of these demonstrates a transition in which the networks proceed from some $f=f_c$ to $f=1$ at a critical voltage or current. 

\section{Cluster Models}

\begin{figure}
	\includegraphics[width=8.6cm]{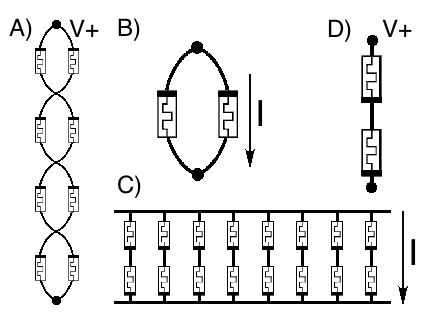}
	\caption{The systems and sub-units considered in the cluster approximations.  As a model of the conducting backbone embedded in the network we consider (A) a series `chain' of memristors in parallel subject to a slowly ramped voltage $V+$ composed of sub-units of (B) pairs of memristors in parallel subject to a slowly ramped current.  As a model of the crack, we consider (C) a parallel strip of memristors in series subject to a slowly ramped current consisting of sub-units of (D) pairs of memristors in series subject to a slowly ramped voltage.
		\label{Cluster_circuits}}
\end{figure}

In order to include the influence of the conducting backbone or crack on the rest of the network and thus the behavior of the network past the transition, we use a cluster approach similar to the Bethe-Kikuchi approximation in equilibrium thermodynamics \cite{Bethe1935, Kikuchi1951}. To this end we replace each memristor in the series chain with two memristors in parallel as in Figure~\ref{Cluster_circuits} A and subject the entire chain to a slowly ramped voltage.  Each pair in the chain is thus subject to a current $I$ slowly raised from zero (See Fig. \ref{Cluster_circuits} B).  If the threshold of each memristor, $t_i$ is drawn independently from a distribution $p(t)$, the probability distribution for the conductance of the pair $G_{\parallel}$ is,
\begin{eqnarray}
p(2G_{off};I) = 2\int_{I/2}^\infty dt_1 \int_{t_1}^\infty dt_2 \, p(t_1) p(t_2) \nonumber\\
p(G_{off} + G_{on}; I) = 2\int_{\frac{G_{off}I}{G_{off} + G_{on}}}^{I/2} dt_1 \int_{t_1}^\infty dt_2 \,p(t_1)p(t_2) \nonumber\\
 + 2\int^{\frac{G_{off}I}{G_{off} + G_{on}}}_0 dt_1 \int_{\frac{G_{off}I}{G_{off} + G_{on}}}^\infty dt_2 \, p(t_1) p(t_2) \nonumber\\
p(2G_{on} ; I) = 2\int_0^{\frac{G_{off}I}{G_{off} + G_{on}}} dt_2 \int_0^{t_2} dt_1 \, p(t_1)p(t_2).\nonumber \\
\end{eqnarray}
A long chain of such pairs in series (Fig. \ref{Cluster_circuits} A), each with conductance $G_{\parallel,i}$ will possess a total conductance
close to
\begin{equation}
\left\langle\frac{1}{G_{chain}}\right\rangle  =  N \left\langle\frac{1}{G_{\parallel}}\right\rangle
\end{equation}
and we may determine the current through the chain self-consistently as the smallest solution of the equation
\begin{equation}
\frac{V/N_x}{I} = \left\langle \frac{1}{G_{\parallel}}\right\rangle
\label{SC_Cluster}
\end{equation}
where the $I$ dependence of $\langle \frac{1}{G_{\parallel}}\rangle$ has entered through the averaging.  Using again the distribution Uniform(0,1), the solution to this equation has been plotted in the insets of Figs. \ref{IV_fig} and \ref{backward_forward} for several values of $G_{on}$. 

Here, we see the finite jump in conductance observed in simulations followed by the gradual switching of the remaining memristors. This is due
to the memristor in the ON state diverting current away from its neighbor.  While a current of only $I = 2$ (in units of $G_{off}V$) is required to switch the first of the pair, a current of $I\approx (G_{on} + G_{off})/G_{off}$ is required to guarantee the switching of the second, which for large values of $G_{on}/G_{off}$ is considerable.

An analogous treatment of the reversed switching process requires replacing each element of the parallel strip of memristors with two elements in series (Fig. \ref{Cluster_circuits} C) subject to a slowly ramped current.  Each pair is then subject to a slowly ramped voltage $V$ (Fig.~\ref{Cluster_circuits} D).  Again drawing the thresholds independently from a distribution $p(t)$, the distribution for the conductance of the pair is
\begin{eqnarray}
p(\frac{G_{on}}{2};V) = 2\int_{G_{on}V/2}^\infty dt_1 \int_{t_1}^\infty dt_2 \, p(t_1) p(t_2) \nonumber\\
p(\frac{G_{on}G_{off}}{G_{on} + G_{off}}; V) = 2\int_{\frac{G_{on}G_{off}V}{G_{off} + G_{on}}}^{G_{on}V/2} dt_1 \int_{t_1}^\infty dt_2 \,p(t_1)p(t_2) \nonumber\\
 + 2\int^{\frac{G_{on}G_{off}V}{G_{off} + G_{on}}}_0 dt_1 \int_{\frac{G_{on}G_{off}V}{G_{off} + G_{on}}}^\infty dt_2 \, p(t_1) p(t_2) \nonumber\\
p(\frac{G_{off}}{2} ; V) = 2\int_0^{\frac{G_{on}G_{off}V}{G_{off} + G_{on}}} dt_2 \int_0^{t_2} dt_1 \, p(t_1)p(t_2).\nonumber \\
\end{eqnarray}

A strip of $N_y$ of these, each with conductance $G_{series,i}$, in parallel (Fig. \ref{Cluster_circuits} C) will have conductance $N_y \langle G_{series}\rangle$.  As discussed above, such a strip embedded within a network will be subject to a current $I$ and thus satisfy
\begin{equation}
 N_y\langle G_{series}\rangle V = I,
 \label{SC_Cluster_r}
 \end{equation}
where the voltage across the strip is determined self-consistently as the smallest solution of the above equation and the voltage dependence of $\langle G_{series}\rangle$ has entered through the averaging over the voltage dependent distribution above.

The two structures considered above are again {\it dual} and while the self-consistency equation may be solved as before,  we instead note that the exchange
\begin{equation}
V\to \frac{IN_y}{N_x G_{on}}, \quad I \to \frac{VG_{off} N_x}{N_y}
\end{equation}
takes the above self-consistency equation, to that of the forward switching process (we consider a square network $N_x = N_y =N$ to avoid dimensional factors)
\begin{equation}
N\langle G_{series}\rangle V = I \to 
\frac{V/N}{I} = \left\langle \frac{1}{G_{\parallel}}\right\rangle
\end{equation}
The reversed switching process thus maps exactly to the forward switching process upon exchanging the roles of the voltage and current and scaling appropriately, as seen in the simulated I-V curves of Figure \ref{backward_forward}.

\section{Mean-Field Dynamics}

Although the avalanche dynamics of mean-field models akin to the random-field Ising model are well known \cite{Sethna1993}, we include here a brief discussion in the interest of completeness.

The mean-field theories considered lead to self-consistency equations of the form
\begin{equation} 
f = \int_0^{h(f) v} p(t) dt
\label{SOC_dynamics}
\end{equation}
where $v$ is an external field and $h(f)$ is some function of the fraction of switched elements $f$. We consider this relation to be initially satisfied and raise the voltage until the next threshold is passed.  This takes $f\to f+\frac{1}{N}$, increasing 
the limit of (\ref{SOC_dynamics}) by $\frac{h'(f)v}{N}$. The probability that $n$ memristors are switched ON
by this increase is given by a Poisson distribution
\begin{equation}
p_n = \frac{\mu^n}{n!} e^{-\mu}, \quad \mu = p(h(f)v) h'(f)v.
\end{equation}
Each of the $n$ memristors will cause a similar increase in the mean-field voltage, and thus will give rise to the same distribution.  Therefore, by switching a single memristor gives rise to a poissonian
branching process.  This may be brought into a more useful form by calculating the total number of
memristors switched in a single branching process, or the avalanche size distribution.  This leads to the Borel distribution
\begin{equation}
p_S = \frac{(\mu S)^{S-1}}{S!} e^{-\mu S}, \quad S=1, 2, ...
\end{equation}
with $\mu= p(h(f)v) h'(f)v$. This distribution has mean and
variance,
\begin{equation}
\langle S \rangle = \frac{1}{1-\mu},\quad\quad\quad \sigma_S^2 = \frac{\mu}{(1-\mu)^3}.
\end{equation}
and therefore the mean-field dynamics give avalanches whose size is determined by the
parameter $\mu$.  

For $\mu < 0$, raising the voltage will cause individual memristors to switch and no avalanches will occur, corresponding to a diffuse regime. For $0<\mu < 1$, the system will display avalanches of finite size according to the Borel distribution.  At $\mu=1$,
the system reaches a critical branching process, at which point the probability of an infinite avalanche begins to grow and the distribution approaches the limiting form $p_S \sim S^{-3/2}$.

The conductance jumps that the system experiences for avalanches in the regime $0<\mu < 1$ should be approximately $G_{net}'(f)\frac{S}{N}$ and thus, for a particular value of the conductance, conductance jumps in that vicinity should follow a Borel distribution.  Such jump avalanche distributions have been well studied both numerically \cite{Shekhawat2011} and in experiment \cite{Sharoni2008} for individual memristive elements but not yet disordered systems consisting of many memristive elements, such as those of Stieg {\it et al.}~\cite{Stieg2014}.

\begin{figure}
\includegraphics[width=8.6cm]{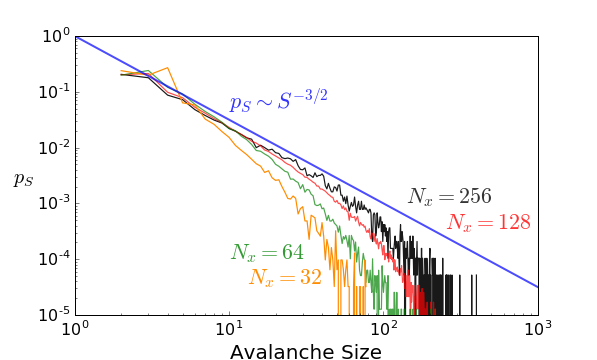}
\caption{Avalanche sizes binned just above the transition for randomly
diluted networks ($p=0.6,\,G_{on}=100$) of size $N_{x/y}=32, 64, 128$ (1000 realizations) and $N_{x/y}=256$ (100 realizations).  As the network size increases,
the avalanche size distribution approaches the asymptotic form $P(s)\sim s^{-3/2}$ given by the
mean-field theory, subject to a finite size cut-off.
\label{Avalanche_fig}}
\end{figure}

In order to confirm whether this scaling law would be accessible in experiments for physically disordered lattices, we have 
simulated randomly diluted lattices (by removing bonds above percolation) without threshold disorder as the
effect of spatial correlations may modify the behavior.  Avalanches were
binned in the region surrounding the peak in the avalanche size for
various sizes of the networks. The histograms produced are plotted in
Figure \ref{Avalanche_fig}. As the system size increases, the histograms approach
the form predicted by the mean-field theory, although clearly subject to a finite-size
cutoff.

\section{Conclusions} 

We have presented a simple model that captures the behavior of a disordered two-dimensional memristive
network when subject to bias in the adiabatic limit.  As the memristive $G_{on}/G_{off}$ ratio is increased,
the conductivity changes from a smooth function of the applied voltage to
displaying a discontinuous jump as in a {\it first-order phase transition}. Internally, this is due to the formation
of a conducting backbone or crack through the network.  While the I-V curves demonstrate such a jump, the restriction of the transition to a small subset of the network elements moderates its size to a fraction of the network conductivity. Furthermore, the
current diverted from the rest of the network extends the voltage range of the remaining memristors, maintaining
the voltage range of the network.  

The $G_{on}\leftrightarrow G_{off}$ processes are connected by a {\it duality} that maps the hysteresis curves of a voltage-controlled network to those of a current-controlled network in the opposite polarity.  A cluster approximation duplicates this behavior and reveals that, in order to fully transition the network, elements of the backbone will need to carry currents a factor $G_{on}/G_{off}$ larger than their neighbors.  As filament-type memristive devices have both large $G_{on}/G_{off}$ ratios and are sensitive to the maximum current through them, this may limit the operating voltages of computational devices manufactured from memristors to the neighborhood of the transition.  Fortunately, this seems to be the region in which the dynamics of the networks carry the greatest promise for the design of computational devices, as seen in maze and shortest path solvers \cite{Pershin2013} where the transition may correspond to the solution of an optimization problem.  We hope this work will provide a foundation to extend the understanding of these networks to the non-adiabatic regime in which their behavior may be substantially more complex and interesting.

\begin{acknowledgments}
We thank G. Pruessner, S. Peotta, F. Caravelli, and F. Traversa for useful discussions and acknowledge partial support from the Center for Memory Recording Research at UCSD.
\end{acknowledgments}

\bibliography{Advancement_PT_Memnets-5}

\begin{thebibliography}{30}%
\makeatletter
\providecommand \@ifxundefined [1]{%
 \@ifx{#1\undefined}
}%
\providecommand \@ifnum [1]{%
 \ifnum #1\expandafter \@firstoftwo
 \else \expandafter \@secondoftwo
 \fi
}%
\providecommand \@ifx [1]{%
 \ifx #1\expandafter \@firstoftwo
 \else \expandafter \@secondoftwo
 \fi
}%
\providecommand \natexlab [1]{#1}%
\providecommand \enquote  [1]{``#1''}%
\providecommand \bibnamefont  [1]{#1}%
\providecommand \bibfnamefont [1]{#1}%
\providecommand \citenamefont [1]{#1}%
\providecommand \href@noop [0]{\@secondoftwo}%
\providecommand \href [0]{\begingroup \@sanitize@url \@href}%
\providecommand \@href[1]{\@@startlink{#1}\@@href}%
\providecommand \@@href[1]{\endgroup#1\@@endlink}%
\providecommand \@sanitize@url [0]{\catcode `\\12\catcode `\$12\catcode
  `\&12\catcode `\#12\catcode `\^12\catcode `\_12\catcode `\%12\relax}%
\providecommand \@@startlink[1]{}%
\providecommand \@@endlink[0]{}%
\providecommand \url  [0]{\begingroup\@sanitize@url \@url }%
\providecommand \@url [1]{\endgroup\@href {#1}{\urlprefix }}%
\providecommand \urlprefix  [0]{URL }%
\providecommand \Eprint [0]{\href }%
\providecommand \doibase [0]{http://dx.doi.org/}%
\providecommand \selectlanguage [0]{\@gobble}%
\providecommand \bibinfo  [0]{\@secondoftwo}%
\providecommand \bibfield  [0]{\@secondoftwo}%
\providecommand \translation [1]{[#1]}%
\providecommand \BibitemOpen [0]{}%
\providecommand \bibitemStop [0]{}%
\providecommand \bibitemNoStop [0]{.\EOS\space}%
\providecommand \EOS [0]{\spacefactor3000\relax}%
\providecommand \BibitemShut  [1]{\csname bibitem#1\endcsname}%
\let\auto@bib@innerbib\@empty
\bibitem [{\citenamefont {Pershin}\ and\ \citenamefont {{Di
  Ventra}}(2011)}]{Pershin2011}%
  \BibitemOpen
  \bibfield  {author} {\bibinfo {author} {\bibfnamefont {Y.~V.}\ \bibnamefont
  {Pershin}}\ and\ \bibinfo {author} {\bibfnamefont {M.}~\bibnamefont {{Di
  Ventra}}},\ }\href {\doibase 10.1080/00018732.2010.544961} {\bibfield
  {journal} {\bibinfo  {journal} {Adv. Phys.}\ }\textbf {\bibinfo {volume}
  {60}},\ \bibinfo {pages} {145} (\bibinfo {year} {2011})}\BibitemShut
  {NoStop}%
\bibitem [{\citenamefont {Hasegawa}\ \emph {et~al.}(2010)\citenamefont
  {Hasegawa}, \citenamefont {Ohno}, \citenamefont {Terabe}, \citenamefont
  {Tsuruoka}, \citenamefont {Nakayama}, \citenamefont {Gimzewski},\ and\
  \citenamefont {Aono}}]{Hasegawa2010}%
  \BibitemOpen
  \bibfield  {author} {\bibinfo {author} {\bibfnamefont {T.}~\bibnamefont
  {Hasegawa}}, \bibinfo {author} {\bibfnamefont {T.}~\bibnamefont {Ohno}},
  \bibinfo {author} {\bibfnamefont {K.}~\bibnamefont {Terabe}}, \bibinfo
  {author} {\bibfnamefont {T.}~\bibnamefont {Tsuruoka}}, \bibinfo {author}
  {\bibfnamefont {T.}~\bibnamefont {Nakayama}}, \bibinfo {author}
  {\bibfnamefont {J.~K.}\ \bibnamefont {Gimzewski}}, \ and\ \bibinfo {author}
  {\bibfnamefont {M.}~\bibnamefont {Aono}},\ }\href {\doibase
  10.1002/adma.200903680} {\bibfield  {journal} {\bibinfo  {journal} {Adv.
  Mater.}\ }\textbf {\bibinfo {volume} {22}},\ \bibinfo {pages} {1831}
  (\bibinfo {year} {2010})}\BibitemShut {NoStop}%
\bibitem [{\citenamefont {Ohno}\ \emph {et~al.}(2011)\citenamefont {Ohno},
  \citenamefont {Hasegawa}, \citenamefont {Tsuruoka}, \citenamefont {Terabe},
  \citenamefont {Gimzewski},\ and\ \citenamefont {Aono}}]{Ohno2011}%
  \BibitemOpen
  \bibfield  {author} {\bibinfo {author} {\bibfnamefont {T.}~\bibnamefont
  {Ohno}}, \bibinfo {author} {\bibfnamefont {T.}~\bibnamefont {Hasegawa}},
  \bibinfo {author} {\bibfnamefont {T.}~\bibnamefont {Tsuruoka}}, \bibinfo
  {author} {\bibfnamefont {K.}~\bibnamefont {Terabe}}, \bibinfo {author}
  {\bibfnamefont {J.~K.}\ \bibnamefont {Gimzewski}}, \ and\ \bibinfo {author}
  {\bibfnamefont {M.}~\bibnamefont {Aono}},\ }\href {\doibase 10.1038/nmat3054}
  {\bibfield  {journal} {\bibinfo  {journal} {Nat. Mater.}\ }\textbf {\bibinfo
  {volume} {10}},\ \bibinfo {pages} {591} (\bibinfo {year} {2011})}\BibitemShut
  {NoStop}%
\bibitem [{\citenamefont {Chialvo}(2010)}]{Chialvo2010}%
  \BibitemOpen
  \bibfield  {author} {\bibinfo {author} {\bibfnamefont {D.~R.}\ \bibnamefont
  {Chialvo}},\ }\href {\doibase 10.1038/nphys1803} {\bibfield  {journal}
  {\bibinfo  {journal} {Nat. Phys.}\ }\textbf {\bibinfo {volume} {6}},\
  \bibinfo {pages} {744} (\bibinfo {year} {2010})}\BibitemShut {NoStop}%
\bibitem [{\citenamefont {Avizienis}\ \emph {et~al.}(2012)\citenamefont
  {Avizienis}, \citenamefont {Sillin}, \citenamefont {Martin-Olmos},
  \citenamefont {Shieh}, \citenamefont {Aono}, \citenamefont {Stieg},\ and\
  \citenamefont {Gimzewski}}]{Avizienis2012}%
  \BibitemOpen
  \bibfield  {author} {\bibinfo {author} {\bibfnamefont {A.~V.}\ \bibnamefont
  {Avizienis}}, \bibinfo {author} {\bibfnamefont {H.~O.}\ \bibnamefont
  {Sillin}}, \bibinfo {author} {\bibfnamefont {C.}~\bibnamefont
  {Martin-Olmos}}, \bibinfo {author} {\bibfnamefont {H.~H.}\ \bibnamefont
  {Shieh}}, \bibinfo {author} {\bibfnamefont {M.}~\bibnamefont {Aono}},
  \bibinfo {author} {\bibfnamefont {A.~Z.}\ \bibnamefont {Stieg}}, \ and\
  \bibinfo {author} {\bibfnamefont {J.~K.}\ \bibnamefont {Gimzewski}},\ }\href
  {\doibase 10.1371/journal.pone.0042772} {\bibfield  {journal} {\bibinfo
  {journal} {PLoS One}\ }\textbf {\bibinfo {volume} {7}},\ \bibinfo {pages}
  {e42772} (\bibinfo {year} {2012})}\BibitemShut {NoStop}%
\bibitem [{\citenamefont {Stieg}\ and\ \citenamefont
  {Avizienis}(2014)}]{Stieg2014}%
  \BibitemOpen
  \bibfield  {author} {\bibinfo {author} {\bibfnamefont {A.}~\bibnamefont
  {Stieg}}\ and\ \bibinfo {author} {\bibfnamefont {A.}~\bibnamefont
  {Avizienis}},\ }\href {http://iopscience.iop.org/1347-4065/53/1S/01AA02}
  {\bibfield  {journal} {\bibinfo  {journal} {Jpn. J. Appl. Phys.}\ }\textbf
  {\bibinfo {volume} {02}} (\bibinfo {year} {2014})}\BibitemShut {NoStop}%
\bibitem [{\citenamefont {Ventra}\ and\ \citenamefont
  {Pershin}(2013)}]{Ventra2013}%
  \BibitemOpen
  \bibfield  {author} {\bibinfo {author} {\bibfnamefont {M.~D.}\ \bibnamefont
  {Ventra}}\ and\ \bibinfo {author} {\bibfnamefont {Y.}~\bibnamefont
  {Pershin}},\ }\href
  {http://www.nature.com/nphys/journal/v9/n4/full/nphys2566.html} {\bibfield
  {journal} {\bibinfo  {journal} {Nat. Phys.}\ }\textbf {\bibinfo {volume}
  {9}},\ \bibinfo {pages} {200} (\bibinfo {year} {2013})}\BibitemShut {NoStop}%
\bibitem [{\citenamefont {Pershin}\ and\ \citenamefont {{Di
  Ventra}}(2013)}]{Pershin2013}%
  \BibitemOpen
  \bibfield  {author} {\bibinfo {author} {\bibfnamefont {Y.~V.}\ \bibnamefont
  {Pershin}}\ and\ \bibinfo {author} {\bibfnamefont {M.}~\bibnamefont {{Di
  Ventra}}},\ }\href {\doibase 10.1103/PhysRevE.88.013305} {\bibfield
  {journal} {\bibinfo  {journal} {Phys. Rev. E}\ }\textbf {\bibinfo {volume}
  {88}},\ \bibinfo {pages} {013305} (\bibinfo {year} {2013})}\BibitemShut
  {NoStop}%
\bibitem [{\citenamefont {Traversa}\ \emph {et~al.}(2015)\citenamefont
  {Traversa}, \citenamefont {Ramella}, \citenamefont {Bonani},\ and\
  \citenamefont {{Di Ventra}}}]{Traversa2015}%
  \BibitemOpen
  \bibfield  {author} {\bibinfo {author} {\bibfnamefont {F.~L.}\ \bibnamefont
  {Traversa}}, \bibinfo {author} {\bibfnamefont {C.}~\bibnamefont {Ramella}},
  \bibinfo {author} {\bibfnamefont {F.}~\bibnamefont {Bonani}}, \ and\ \bibinfo
  {author} {\bibfnamefont {M.}~\bibnamefont {{Di Ventra}}},\ }\href {\doibase
  10.1126/sciadv.1500031} {\bibfield  {journal} {\bibinfo  {journal} {Sci.
  Adv.}\ }\textbf {\bibinfo {volume} {1}},\ \bibinfo {pages} {e1500031}
  (\bibinfo {year} {2015})}\BibitemShut {NoStop}%
\bibitem [{\citenamefont {Traversa}\ and\ \citenamefont {{Di
  Ventra}}(2015)}]{Traversa15}%
  \BibitemOpen
  \bibfield  {author} {\bibinfo {author} {\bibfnamefont {F.~L.}\ \bibnamefont
  {Traversa}}\ and\ \bibinfo {author} {\bibfnamefont {M.}~\bibnamefont {{Di
  Ventra}}},\ }\href {\doibase 10.1109/TNNLS.2015.2391182} {\bibfield
  {journal} {\bibinfo  {journal} {IEEE Trans. Neural Networks Learn. Syst.}\
  }\textbf {\bibinfo {volume} {26}},\ \bibinfo {pages} {1} (\bibinfo {year}
  {2015})}\BibitemShut {NoStop}%
\bibitem [{\citenamefont {Langton}(1990)}]{Langton1990}%
  \BibitemOpen
  \bibfield  {author} {\bibinfo {author} {\bibfnamefont {C.}~\bibnamefont
  {Langton}},\ }\href
  {http://www.sciencedirect.com/science/article/pii/016727899090064V}
  {\bibfield  {journal} {\bibinfo  {journal} {Phys. D Nonlinear Phenom.}\
  }\textbf {\bibinfo {volume} {42}},\ \bibinfo {pages} {12} (\bibinfo {year}
  {1990})}\BibitemShut {NoStop}%
\bibitem [{\citenamefont {Stieg}\ and\ \citenamefont
  {Avizienis}(2012)}]{Stieg2012}%
  \BibitemOpen
  \bibfield  {author} {\bibinfo {author} {\bibfnamefont {A.}~\bibnamefont
  {Stieg}}\ and\ \bibinfo {author} {\bibfnamefont {A.}~\bibnamefont
  {Avizienis}},\ }\href {\doibase 10.1002/adma.201103053} {\bibfield  {journal}
  {\bibinfo  {journal} {Adv. Mater.}\ }\textbf {\bibinfo {volume} {24}},\
  \bibinfo {pages} {286} (\bibinfo {year} {2012})}\BibitemShut {NoStop}%
\bibitem [{\citenamefont {{Nedaaee Oskoee}}\ and\ \citenamefont
  {Sahimi}(2011)}]{NedaaeeOskoee2011}%
  \BibitemOpen
  \bibfield  {author} {\bibinfo {author} {\bibfnamefont {E.}~\bibnamefont
  {{Nedaaee Oskoee}}}\ and\ \bibinfo {author} {\bibfnamefont {M.}~\bibnamefont
  {Sahimi}},\ }\href {\doibase 10.1103/PhysRevE.83.031105} {\bibfield
  {journal} {\bibinfo  {journal} {Phys. Rev. E}\ }\textbf {\bibinfo {volume}
  {83}},\ \bibinfo {pages} {031105} (\bibinfo {year} {2011})}\BibitemShut
  {NoStop}%
\bibitem [{\citenamefont {Sillin}\ \emph {et~al.}(2013)\citenamefont {Sillin},
  \citenamefont {Aguilera}, \citenamefont {Shieh}, \citenamefont {Avizienis},
  \citenamefont {Aono}, \citenamefont {Stieg},\ and\ \citenamefont
  {Gimzewski}}]{Sillin2013}%
  \BibitemOpen
  \bibfield  {author} {\bibinfo {author} {\bibfnamefont {H.~O.}\ \bibnamefont
  {Sillin}}, \bibinfo {author} {\bibfnamefont {R.}~\bibnamefont {Aguilera}},
  \bibinfo {author} {\bibfnamefont {H.-H.}\ \bibnamefont {Shieh}}, \bibinfo
  {author} {\bibfnamefont {A.~V.}\ \bibnamefont {Avizienis}}, \bibinfo {author}
  {\bibfnamefont {M.}~\bibnamefont {Aono}}, \bibinfo {author} {\bibfnamefont
  {A.~Z.}\ \bibnamefont {Stieg}}, \ and\ \bibinfo {author} {\bibfnamefont
  {J.~K.}\ \bibnamefont {Gimzewski}},\ }\href {\doibase
  10.1088/0957-4484/24/38/384004} {\bibfield  {journal} {\bibinfo  {journal}
  {Nanotechnology}\ }\textbf {\bibinfo {volume} {24}},\ \bibinfo {pages}
  {384004} (\bibinfo {year} {2013})}\BibitemShut {NoStop}%
\bibitem [{\citenamefont {{Di Ventra}}\ and\ \citenamefont
  {Pershin}(2013)}]{diVentra2013}%
  \BibitemOpen
  \bibfield  {author} {\bibinfo {author} {\bibfnamefont {M.}~\bibnamefont {{Di
  Ventra}}}\ and\ \bibinfo {author} {\bibfnamefont {Y.~V.}\ \bibnamefont
  {Pershin}},\ }\href {\doibase 10.1088/0957-4484/24/25/255201} {\bibfield
  {journal} {\bibinfo  {journal} {Nanotechnology}\ }\textbf {\bibinfo {volume}
  {24}},\ \bibinfo {pages} {255201} (\bibinfo {year} {2013})}\BibitemShut
  {NoStop}%
\bibitem [{\citenamefont {Inoue}\ \emph {et~al.}(2008)\citenamefont {Inoue},
  \citenamefont {Yasuda}, \citenamefont {Akinaga},\ and\ \citenamefont
  {Takagi}}]{Inoue2008}%
  \BibitemOpen
  \bibfield  {author} {\bibinfo {author} {\bibfnamefont {I.~H.}\ \bibnamefont
  {Inoue}}, \bibinfo {author} {\bibfnamefont {S.}~\bibnamefont {Yasuda}},
  \bibinfo {author} {\bibfnamefont {H.}~\bibnamefont {Akinaga}}, \ and\
  \bibinfo {author} {\bibfnamefont {H.}~\bibnamefont {Takagi}},\ }\href
  {\doibase 10.1103/PhysRevB.77.035105} {\bibfield  {journal} {\bibinfo
  {journal} {Phys. Rev. B - Condens. Matter Mater. Phys.}\ }\textbf {\bibinfo
  {volume} {77}},\ \bibinfo {pages} {1} (\bibinfo {year} {2008})}\BibitemShut
  {NoStop}%
\bibitem [{\citenamefont {Sarswat}\ \emph {et~al.}(2015)\citenamefont
  {Sarswat}, \citenamefont {Smith}, \citenamefont {Free},\ and\ \citenamefont
  {Misra}}]{Sarswat2015}%
  \BibitemOpen
  \bibfield  {author} {\bibinfo {author} {\bibfnamefont {P.~K.}\ \bibnamefont
  {Sarswat}}, \bibinfo {author} {\bibfnamefont {Y.~R.}\ \bibnamefont {Smith}},
  \bibinfo {author} {\bibfnamefont {M.~L.}\ \bibnamefont {Free}}, \ and\
  \bibinfo {author} {\bibfnamefont {M.}~\bibnamefont {Misra}},\ }\href
  {\doibase 10.1149/2.0071508jss} {\bibfield  {journal} {\bibinfo  {journal}
  {ECS J. Solid State Sci. Technol.}\ }\textbf {\bibinfo {volume} {4}},\
  \bibinfo {pages} {Q83} (\bibinfo {year} {2015})}\BibitemShut {NoStop}%
\bibitem [{\citenamefont {Sillin}\ \emph {et~al.}(2014)\citenamefont {Sillin},
  \citenamefont {Sandouk}, \citenamefont {Avizienis}, \citenamefont {Aono},
  \citenamefont {Stieg},\ and\ \citenamefont {Gimzewski}}]{Sillin2014}%
  \BibitemOpen
  \bibfield  {author} {\bibinfo {author} {\bibfnamefont {H.~O.}\ \bibnamefont
  {Sillin}}, \bibinfo {author} {\bibfnamefont {E.~J.}\ \bibnamefont {Sandouk}},
  \bibinfo {author} {\bibfnamefont {A.~V.}\ \bibnamefont {Avizienis}}, \bibinfo
  {author} {\bibfnamefont {M.}~\bibnamefont {Aono}}, \bibinfo {author}
  {\bibfnamefont {A.~Z.}\ \bibnamefont {Stieg}}, \ and\ \bibinfo {author}
  {\bibfnamefont {J.~K.}\ \bibnamefont {Gimzewski}},\ }\href
  {http://www.ncbi.nlm.nih.gov/pubmed/24734692} {\bibfield  {journal} {\bibinfo
   {journal} {J. Nanosci. Nanotechnol.}\ }\textbf {\bibinfo {volume} {14}},\
  \bibinfo {pages} {2792} (\bibinfo {year} {2014})}\BibitemShut {NoStop}%
\bibitem [{\citenamefont {Xu}\ \emph {et~al.}(2010)\citenamefont {Xu},
  \citenamefont {Bando}, \citenamefont {Wang}, \citenamefont {Bai},\ and\
  \citenamefont {Golberg}}]{Xu2010}%
  \BibitemOpen
  \bibfield  {author} {\bibinfo {author} {\bibfnamefont {Z.}~\bibnamefont
  {Xu}}, \bibinfo {author} {\bibfnamefont {Y.}~\bibnamefont {Bando}}, \bibinfo
  {author} {\bibfnamefont {W.}~\bibnamefont {Wang}}, \bibinfo {author}
  {\bibfnamefont {X.}~\bibnamefont {Bai}}, \ and\ \bibinfo {author}
  {\bibfnamefont {D.}~\bibnamefont {Golberg}},\ }\href
  {http://pubs.acs.org/doi/abs/10.1021/nn100483a} {\bibfield  {journal}
  {\bibinfo  {journal} {ACS Nano}\ }\textbf {\bibinfo {volume} {4}},\ \bibinfo
  {pages} {2515} (\bibinfo {year} {2010})}\BibitemShut {NoStop}%
\bibitem [{\citenamefont {{Di Ventra}}(2008)}]{di2008electrical}%
  \BibitemOpen
  \bibfield  {author} {\bibinfo {author} {\bibfnamefont {M.}~\bibnamefont {{Di
  Ventra}}},\ }\href@noop {} {\emph {\bibinfo {title} {{Electrical transport in
  nanoscale systems}}}},\ Vol.~\bibinfo {volume} {14}\ (\bibinfo  {publisher}
  {Cambridge University Press Cambridge},\ \bibinfo {year} {2008})\BibitemShut
  {NoStop}%
\bibitem [{\citenamefont {Hasegawa}\ \emph {et~al.}(2012)\citenamefont
  {Hasegawa}, \citenamefont {Terabe}, \citenamefont {Tsuruoka},\ and\
  \citenamefont {Aono}}]{Hasegawa2012}%
  \BibitemOpen
  \bibfield  {author} {\bibinfo {author} {\bibfnamefont {T.}~\bibnamefont
  {Hasegawa}}, \bibinfo {author} {\bibfnamefont {K.}~\bibnamefont {Terabe}},
  \bibinfo {author} {\bibfnamefont {T.}~\bibnamefont {Tsuruoka}}, \ and\
  \bibinfo {author} {\bibfnamefont {M.}~\bibnamefont {Aono}},\ }\href {\doibase
  10.1002/adma.201102597} {\bibfield  {journal} {\bibinfo  {journal} {Adv.
  Mater.}\ }\textbf {\bibinfo {volume} {24}},\ \bibinfo {pages} {252} (\bibinfo
  {year} {2012})}\BibitemShut {NoStop}%
\bibitem [{\citenamefont {Kahng}\ \emph {et~al.}(1988)\citenamefont {Kahng},
  \citenamefont {Batrouni}, \citenamefont {Redner}, \citenamefont {{De
  Arcangelis}},\ and\ \citenamefont {Herrmann}}]{kahng1988electrical}%
  \BibitemOpen
  \bibfield  {author} {\bibinfo {author} {\bibfnamefont {B.}~\bibnamefont
  {Kahng}}, \bibinfo {author} {\bibfnamefont {G.~G.}\ \bibnamefont {Batrouni}},
  \bibinfo {author} {\bibfnamefont {S.}~\bibnamefont {Redner}}, \bibinfo
  {author} {\bibfnamefont {L.}~\bibnamefont {{De Arcangelis}}}, \ and\ \bibinfo
  {author} {\bibfnamefont {H.~J.}\ \bibnamefont {Herrmann}},\ }\href@noop {}
  {\bibfield  {journal} {\bibinfo  {journal} {Phys. Rev. B}\ }\textbf {\bibinfo
  {volume} {37}},\ \bibinfo {pages} {7625} (\bibinfo {year}
  {1988})}\BibitemShut {NoStop}%
\bibitem [{\citenamefont {Shekhawat}\ \emph {et~al.}(2011)\citenamefont
  {Shekhawat}, \citenamefont {Papanikolaou}, \citenamefont {Zapperi},\ and\
  \citenamefont {Sethna}}]{Shekhawat2011}%
  \BibitemOpen
  \bibfield  {author} {\bibinfo {author} {\bibfnamefont {A.}~\bibnamefont
  {Shekhawat}}, \bibinfo {author} {\bibfnamefont {S.}~\bibnamefont
  {Papanikolaou}}, \bibinfo {author} {\bibfnamefont {S.}~\bibnamefont
  {Zapperi}}, \ and\ \bibinfo {author} {\bibfnamefont {J.~P.}\ \bibnamefont
  {Sethna}},\ }\href {\doibase 10.1103/PhysRevLett.107.276401} {\bibfield
  {journal} {\bibinfo  {journal} {Phys. Rev. Lett.}\ }\textbf {\bibinfo
  {volume} {107}},\ \bibinfo {pages} {1} (\bibinfo {year} {2011})}\BibitemShut
  {NoStop}%
\bibitem [{\citenamefont {Chae}\ \emph {et~al.}(2008)\citenamefont {Chae},
  \citenamefont {Lee}, \citenamefont {Kim}, \citenamefont {Lee}, \citenamefont
  {Chang}, \citenamefont {Liu}, \citenamefont {Kahng}, \citenamefont {Shin},
  \citenamefont {Kim}, \citenamefont {Jung}, \citenamefont {Seo}, \citenamefont
  {Lee},\ and\ \citenamefont {Noh}}]{Chae2008}%
  \BibitemOpen
  \bibfield  {author} {\bibinfo {author} {\bibfnamefont {S.~C.}\ \bibnamefont
  {Chae}}, \bibinfo {author} {\bibfnamefont {J.~S.}\ \bibnamefont {Lee}},
  \bibinfo {author} {\bibfnamefont {S.}~\bibnamefont {Kim}}, \bibinfo {author}
  {\bibfnamefont {S.~B.}\ \bibnamefont {Lee}}, \bibinfo {author} {\bibfnamefont
  {S.~H.}\ \bibnamefont {Chang}}, \bibinfo {author} {\bibfnamefont
  {C.}~\bibnamefont {Liu}}, \bibinfo {author} {\bibfnamefont {B.}~\bibnamefont
  {Kahng}}, \bibinfo {author} {\bibfnamefont {H.}~\bibnamefont {Shin}},
  \bibinfo {author} {\bibfnamefont {D.~W.}\ \bibnamefont {Kim}}, \bibinfo
  {author} {\bibfnamefont {C.~U.}\ \bibnamefont {Jung}}, \bibinfo {author}
  {\bibfnamefont {S.}~\bibnamefont {Seo}}, \bibinfo {author} {\bibfnamefont
  {M.~J.}\ \bibnamefont {Lee}}, \ and\ \bibinfo {author} {\bibfnamefont
  {T.~W.}\ \bibnamefont {Noh}},\ }\href {\doibase 10.1002/adma.200702024}
  {\bibfield  {journal} {\bibinfo  {journal} {Adv. Mater.}\ }\textbf {\bibinfo
  {volume} {20}},\ \bibinfo {pages} {1154} (\bibinfo {year}
  {2008})}\BibitemShut {NoStop}%
\bibitem [{\citenamefont {Iyer}(1985)}]{Iyer1985}%
  \BibitemOpen
  \bibfield  {author} {\bibinfo {author} {\bibfnamefont {T.}~\bibnamefont
  {Iyer}},\ }\href@noop {} {\emph {\bibinfo {title} {{Circuit Theory}}}}\
  (\bibinfo  {publisher} {Tata McGraw-Hill},\ \bibinfo {address} {New Delhi},\
  \bibinfo {year} {1985})\ pp.\ \bibinfo {pages} {206--212}\BibitemShut
  {NoStop}%
\bibitem [{\citenamefont {Zapperi}\ \emph {et~al.}(1999)\citenamefont
  {Zapperi}, \citenamefont {Ray}, \citenamefont {Stanley},\ and\ \citenamefont
  {Vespignani}}]{Zapperi1999}%
  \BibitemOpen
  \bibfield  {author} {\bibinfo {author} {\bibfnamefont {S.}~\bibnamefont
  {Zapperi}}, \bibinfo {author} {\bibfnamefont {P.}~\bibnamefont {Ray}},
  \bibinfo {author} {\bibfnamefont {H.~E.}\ \bibnamefont {Stanley}}, \ and\
  \bibinfo {author} {\bibfnamefont {A.}~\bibnamefont {Vespignani}},\ }\href
  {\doibase 10.1103/PhysRevE.59.5049} {\bibfield  {journal} {\bibinfo
  {journal} {Phys. Rev. E}\ }\textbf {\bibinfo {volume} {59}},\ \bibinfo
  {pages} {5049} (\bibinfo {year} {1999})}\BibitemShut {NoStop}%
\bibitem [{\citenamefont {Bethe}(1935)}]{Bethe1935}%
  \BibitemOpen
  \bibfield  {author} {\bibinfo {author} {\bibfnamefont {H.~A.}\ \bibnamefont
  {Bethe}},\ }\href@noop {} {\bibfield  {journal} {\bibinfo  {journal} {Proc.
  R. Soc. London A}\ }\textbf {\bibinfo {volume} {150}},\ \bibinfo {pages}
  {552} (\bibinfo {year} {1935})}\BibitemShut {NoStop}%
\bibitem [{\citenamefont {Kikuchi}(1951)}]{Kikuchi1951}%
  \BibitemOpen
  \bibfield  {author} {\bibinfo {author} {\bibfnamefont {R.}~\bibnamefont
  {Kikuchi}},\ }\href {\doibase 10.1017/CBO9781107415324.004} {\bibfield
  {journal} {\bibinfo  {journal} {Phys. Rev.}\ }\textbf {\bibinfo {volume}
  {81}},\ \bibinfo {pages} {988} (\bibinfo {year} {1951})}\BibitemShut
  {NoStop}%
\bibitem [{\citenamefont {Sethna}\ \emph {et~al.}(1993)\citenamefont {Sethna},
  \citenamefont {Dahmen}, \citenamefont {Kartha}, \citenamefont {Krumhansl},
  \citenamefont {Roberts},\ and\ \citenamefont {Shore}}]{Sethna1993}%
  \BibitemOpen
  \bibfield  {author} {\bibinfo {author} {\bibfnamefont {J.~P.}\ \bibnamefont
  {Sethna}}, \bibinfo {author} {\bibfnamefont {K.}~\bibnamefont {Dahmen}},
  \bibinfo {author} {\bibfnamefont {S.}~\bibnamefont {Kartha}}, \bibinfo
  {author} {\bibfnamefont {J.~A.}\ \bibnamefont {Krumhansl}}, \bibinfo {author}
  {\bibfnamefont {B.~W.}\ \bibnamefont {Roberts}}, \ and\ \bibinfo {author}
  {\bibfnamefont {J.~D.}\ \bibnamefont {Shore}},\ }\href {\doibase
  10.1103/PhysRevLett.70.3347} {\bibfield  {journal} {\bibinfo  {journal}
  {Phys. Rev. Lett.}\ }\textbf {\bibinfo {volume} {70}},\ \bibinfo {pages}
  {3347} (\bibinfo {year} {1993})}\BibitemShut {NoStop}%
\bibitem [{\citenamefont {Sharoni}\ \emph {et~al.}(2008)\citenamefont
  {Sharoni}, \citenamefont {Ram{\'{i}}rez},\ and\ \citenamefont
  {Schuller}}]{Sharoni2008}%
  \BibitemOpen
  \bibfield  {author} {\bibinfo {author} {\bibfnamefont {A.}~\bibnamefont
  {Sharoni}}, \bibinfo {author} {\bibfnamefont {J.~G.}\ \bibnamefont
  {Ram{\'{i}}rez}}, \ and\ \bibinfo {author} {\bibfnamefont {I.~K.}\
  \bibnamefont {Schuller}},\ }\href {\doibase 10.1103/PhysRevLett.101.026404}
  {\bibfield  {journal} {\bibinfo  {journal} {Phys. Rev. Lett.}\ }\textbf
  {\bibinfo {volume} {101}},\ \bibinfo {pages} {4} (\bibinfo {year}
  {2008})}\BibitemShut {NoStop}%
\end{thebibliography}%

\end{document}